\documentclass[12pt,preprint]{emulateapj}

\shorttitle{Pits formation from volatile outgassing on 67P/Churyumov-Gerasimenko}
\shortauthors{Mousis et al.}

\usepackage{etex}
\usepackage{m-pictex, m-ch-en}
\usepackage {amsmath}
\usepackage[squaren, Gray, cdot]{SIunits}
\usepackage{color}

\begin{document}

%% LaTeX will automatically break titles if they run longer than
%% one line. However, you may use \\ to force a line break if
%% you desire.

\title{Pits formation from volatile outgassing on 67P/Churyumov-Gerasimenko}

%% Use \author, \affil, and the \and command to format
%% author and affiliation information.
%% Note that \email has replaced the old \authoremail command
%% from AASTeX v4.0. You can use \email to mark an email address
%% anywhere in the paper, not just in the front matter.
%% As in the title, use \\ to force line breaks.

\author{O. Mousis\altaffilmark{1}, A. Guilbert-Lepoutre\altaffilmark{2}, B. Brugger\altaffilmark{1}, L. Jorda\altaffilmark{1}, J. S. Kargel\altaffilmark{3}, A. Bouquet\altaffilmark{4}, A.-T. Auger\altaffilmark{1}, P. Lamy\altaffilmark{1}, P. Vernazza\altaffilmark{1}, N. Thomas\altaffilmark{5}, H. Sierks\altaffilmark{6}}

%% Notice that each of these authors has alternate affiliations, which
%% are identified by the \altaffilmark after each name.  Specify alternate
%% affiliation information with \altaffiltext, with one command per each
%% affiliation.

\altaffiltext{1}{Aix Marseille Universit{\'e}, CNRS, LAM (Laboratoire d'Astrophysique de Marseille) UMR 7326, 13388, Marseille, France {\tt olivier.mousis@lam.fr}}
\altaffiltext{2}{Institut UTINAM, UMR 6213 CNRS-Universit\'e de Franche Comt\'e, Besan\c con, France}
\altaffiltext{3}{Department of Hydrology and Water Resources, University of Arizona, Tucson, USA}
\altaffiltext{4}{Department of Physics and Astronomy, University of Texas at San Antonio, San Antonio, Texas USA; Space Science and Engineering Division, Southwest Research Institute, San Antonio, Texas USA}
\altaffiltext{5}{Physikalisches Institut, Sidlerstrasse 5, University of Bern, CH-3012 Bern, Switzerland}
\altaffiltext{6}{Max-Planck-Institut f\"ur Sonnensystemforschung, 37077 G\"ottingen, Germany}

%% Mark off your abstract in the ``abstract'' environment. In the manuscript
%% style, abstract will output a Received/Accepted line after the
%% title and affiliation information. No date will appear since the author
%% does not have this information. The dates will be filled in by the
%% editorial office after submission.

\begin{abstract}
We investigate the thermal evolution of comet 67P/Churyumov-Gerasimenko's subsurface in the Seth$\_$01 region, where active pits have been observed by the ESA/Rosetta mission. {Our simulations show that clathrate destabilization and amorphous ice crystallization can occur at depths corresponding to those of the observed pits} in a timescale shorter than 67P/Churyumov-Gerasimenko's lifetime in the comet's activity zone in the inner solar system. {Sublimation of crystalline ice down to such depths is possible only in the absence of a dust mantle}, which requires the presence of dust grains in the matrix small enough to be dragged out by gas from the pores. Our results are consistent with both pits formation via sinkholes or subsequent to outbursts, the dominant process depending on the status of the subsurface porosity. A sealed dust mantle would favor episodic and disruptive outgassing as a result of an increasing gas pressure in the pores, while a high porosity should allow the formation of large voids in the subsurface due to the continuous escape of volatiles. We finally conclude that the subsurface of 67P/Churyumov-Gerasimenko is not uniform at a spatial scale of $\sim$100--200~m. 
\end{abstract}


\section{Introduction}

\begin{figure*}[h]
\begin{center}
%\resizebox{\hsize}{!}{\includegraphics[angle=-90]{figure1.pdf}}
\resizebox{\hsize}{!}{\includegraphics[angle=-90,scale=0.8]{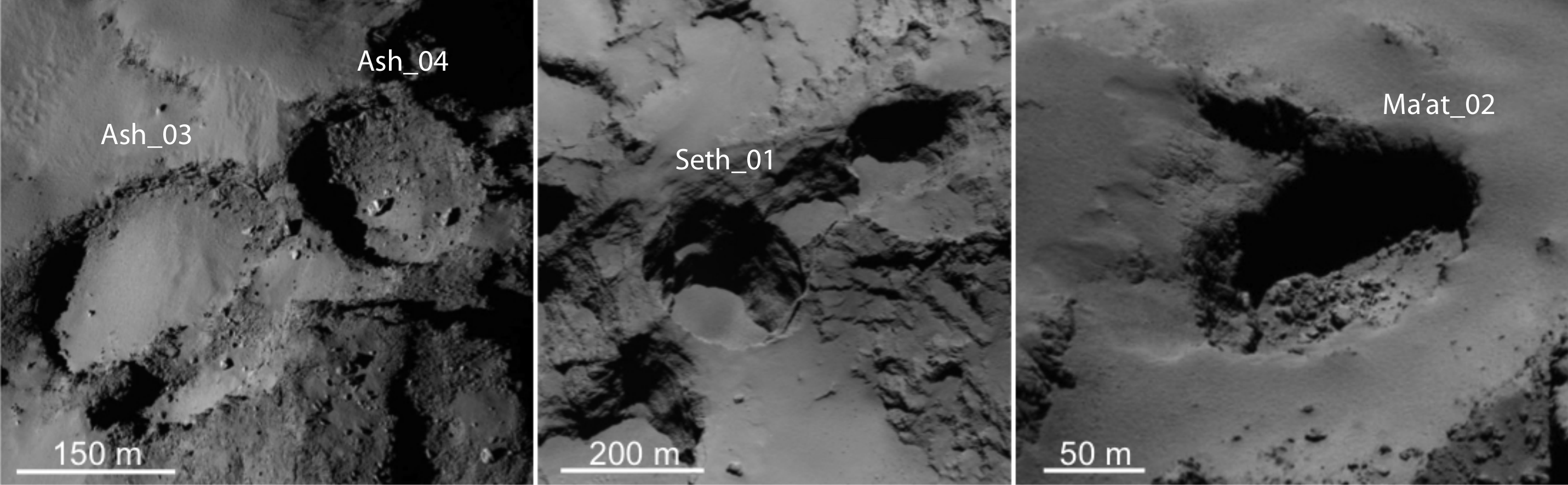}}
\caption{Several examples of pits located on 67P's northern hemisphere and imaged on (from left to right) September 20, 2014 (Ash region), August 28, 2014 (Seth region) and September 12, 2014 (Ma'at region) by the NavCam aboard the ESA/Rosetta spacecraft. Names of pits are indicated on the Figure.}
\label{fig:pits}
\end{center}
\end{figure*}

High resolution images obtained by the OSIRIS instrument and the Navigation Camera (NavCam) onboard the ESA/Rosetta spacecraft have revealed that active circular depressions are a common feature on the surface of comet 67P/Churyumov-Gerasimenko (hereafter 67P, see examples in Fig. \ref{fig:pits}). {Similar depressions or pits likely exist on the surfaces of other comets investigated by previous spacecraft missions. Pits were mapped on 81P/Wild 2 (Brownlee et al. 2004) and 9P/Tempel 1 (Thomas et al. 2013) while their presence has been suspected on 1P/Halley (Keller et al. 1988) and 19P/Borrelly (Solderblom et al. 2002).} The origin of pits on these comets was ascribed to different processes. Belton \& Melosh (2009) proposed that some of the large depressions observed on the surface of 9P/Tempel 1 (9P) might result from the collapse of subsurface cavity's roof, after depletion of a volatile material due to sublimation. They argued that this process could be the result of the transition from amorphous to crystalline water ice, an exothermic process releasing material that could also lead to the local sublimation of surrounding CO and CO$_2$ ices. However, both Belton et al. (2013) and Thomas et al. (2013) ascribed the majority of pits and depressions observed on 9P to the removal of material due to explosive outbursts. 

Vincent et al. (2015) have studied the pits present on 67P's surface, showing that they feature a high depth-to-diameter (d/D) ratio, especially the active ones located in the Seth region, and that they do not display any disturbed material in their surroundings. They investigated several mechanisms for explaining their formation, including impacts, erosion due to insolation, excavation by outbursts, or sinkhole formation resulting from the collapse of a cavity's ceiling. The impact formation hypothesis was discarded because the pits' size distribution is inconsistent with typical impactor size distributions, such that impacts can explain only 4\% of the pits (Belton et al. 2013). Erosion due to insolation would be expected to produce elongated features following {the predominant direction of illumination at perihelion}, which is not observed among 67P's pits. In addition, most of the pits are located in polar night when the nucleus is at perihelion, which severely limits the total energy received in this region. {Based on their estimate that insolation alone would require more than the lifetime of 67P as an active comet (Groussin et al. 2007) to form these pits, Vincent et al. (2015) discarded this process as a possible formation mechanism.} In contrast, they argued that most of the circular depressions could result from sinkhole formation, where the cavity would have been formed after subsurface volatile outgassing. 

In this letter, we examine a series of thermal processes resulting in the local depletion of volatiles, either in the subsurface thus forming cavities, or at the surface leading to an outburst. We note that outbursts have been observed on 67P with the OSIRIS camera between 2014 March 23 and June 24 (Tubiana et al. 2015). We make plausible assumptions on the structure and composition of the material on 67P's subsurface material, and use a thermal evolution model which includes various phase transitions, heat transfer in the ice-dust matrix, and gas diffusion throughout the porous material, based on Marboeuf et al. (2012). We investigate the conditions needed for clathrate destabilization, ice sublimation and amorphous ice crystallization on 67P along its orbital evolution. We show that the mechanisms at the origin of active pits on 67P can be multiple, and we place some constraints on the thermo-physical properties required for forming pits at the surface of this comet.

\section{67P's nucleus model}

In order to evaluate the effect of various thermal processes possibly affecting the surface and subsurface of 67P, we use a quasi-3D thermal evolution model based on physical and numerical schemes described in Marboeuf et al. (2012). At the surface, the 3D description of the illumination conditions (although not accounting for the lateral heat fluxes) has proven to be superior to a 1D description using either a slow or a fast rotator approximation. In the following, we focus on simulating the evolution of the surface and subsurface of the zone where the active pit Seth$\_$01 is located. To do so, we need to parameterize our model so that the spherical approximation used in the numerical scheme can correctly evaluate the illumination conditions in this region. The obliquity and argument of the subsolar meridian at perihelion are therefore calculated from the spin direction, derived from the shape reconstruction of the nucleus (Preusker et al. 2015). The center coordinates and the normal to the surface of pit Seth$\_$01 are calculated from geo-referenced images, such as those presented in Fig.~\ref{fig:pits}, using the stereophotoclinometry shape solution of Jorda et al. (2014). This allows us to calculate the ground geometric parameters, i.e. surface coordinates and normal vector, corresponding to any pixel coordinate of the images. In order to obtain the {\it initial} center coordinates and normal vector of pit Seth$\_$01, we average the coordinates and the normal vector for a set of 45 points, carefully selected around it. 

The model then uses a 1D description of the equations for the internal parts of the nucleus, which is a good first order approximation given the very complex shape and structure revealed by the OSIRIS and NavCam images. We assume that the cometary material is made of a porous mixture of water ice, volatiles and dust. The model describes: i) 1D heat conduction, ii) latent heat exchanges via sublimation and condensation of volatiles, amorphous-crystalline water ice phase transition and clathrate destabilization, iii) gas diffusion in the pores, iv) gas and dust release, and v) dust mantle formation at the surface. Unless otherwise specified, the progressive formation of a porous dust mantle at the surface of the Seth$\_$01 region is the default assumption in our simulations. This is consistent with the ESA/Rosetta observations of the regions where the pits are located, where most of the surface is not active and most probably covered with dust (Thomas et al. 2015). In addition, the asymmetry suggested by the pre- and post-perihelion activity was already recognized as a strong indication that such a crust is indeed present on at least part of the surface (Guilbert-Lepoutre et al. 2014 and references therein). The exact composition is specified in each of the following sections addressing the effect of each thermal process, i.e. clathrate destabilization, sublimation or crystallization. Other thermo-physical parameters are standard, based on previous studies of the thermal evolution of 67P (De Sanctis et al. 2005, Kossacki \& Szutowicz 2006 or Rosenberg \& Prialnik 2009; Marboeuf et al. 2012), and  updated when relevant where the ESA/Rosetta mission has provided new constraints. Table ~\ref{param} summarizes the main parameters of this model. In particular, we choose a thermal inertia of 100 W K$^{-1}$m$^{-2}$s$^{1/2}$. This value is within the 10 to 150 W K$^{-1}$m$^{-2}$s$^{1/2}$ range of thermal inertias measured by the Rosetta/VIRTIS instrument on 67P (Leyrat et al. 2015).

\begin{table}[h]
\centering 
\caption{Modeling parameters for the nucleus}
\begin{tabular}{lcc}
\hline 
\hline
Parameter										& Value						& Reference			\\
\hline
Rotation period	(hr)								& 12.4						& Mottola et al. (2014)	\\
Obliquity ($\degree$)								& 52.3						& 			\\
$\Phi$ ($\degree$)\footnote{Argument of subsolar meridian at perihelion.}	& -111			&			\\
Co-latitude ($\degree$)\footnote{Angle between the normal to the surface and the equatorial plane.}	& 44.1		& \\
Bolometric albedo (\%)							& 1.5							& Fornasier et al. (2015)	\\
Dust-to-ice ratio								& 4							& Rotundi et al. (2014)  	\\
Mean porosity (\%)								& 76							& 			\\
Density (kg/m$^3$)								& 510 						& Jorda et al. (2014)		\\
I (W K$^{-1}$m$^{-2}$s$^{1/2}$)\footnote{Thermal inertia.} 	& 100						& Leyrat et al. (2015)		\\
Assumed CO/CO$_2$ ratio\footnote{Value observed in the northern hemisphere, where the pits showed in Fig. \ref{fig:pits} are located.}	& 1.08	& Le Roy et al. (2015)	\\
\hline
\end{tabular}
\label{param}
\end{table}

\section{Devolatilization processes}

\subsection{Clathrate destabilization}

\begin{figure}
\begin{center}
\includegraphics[angle=0,scale=0.5]{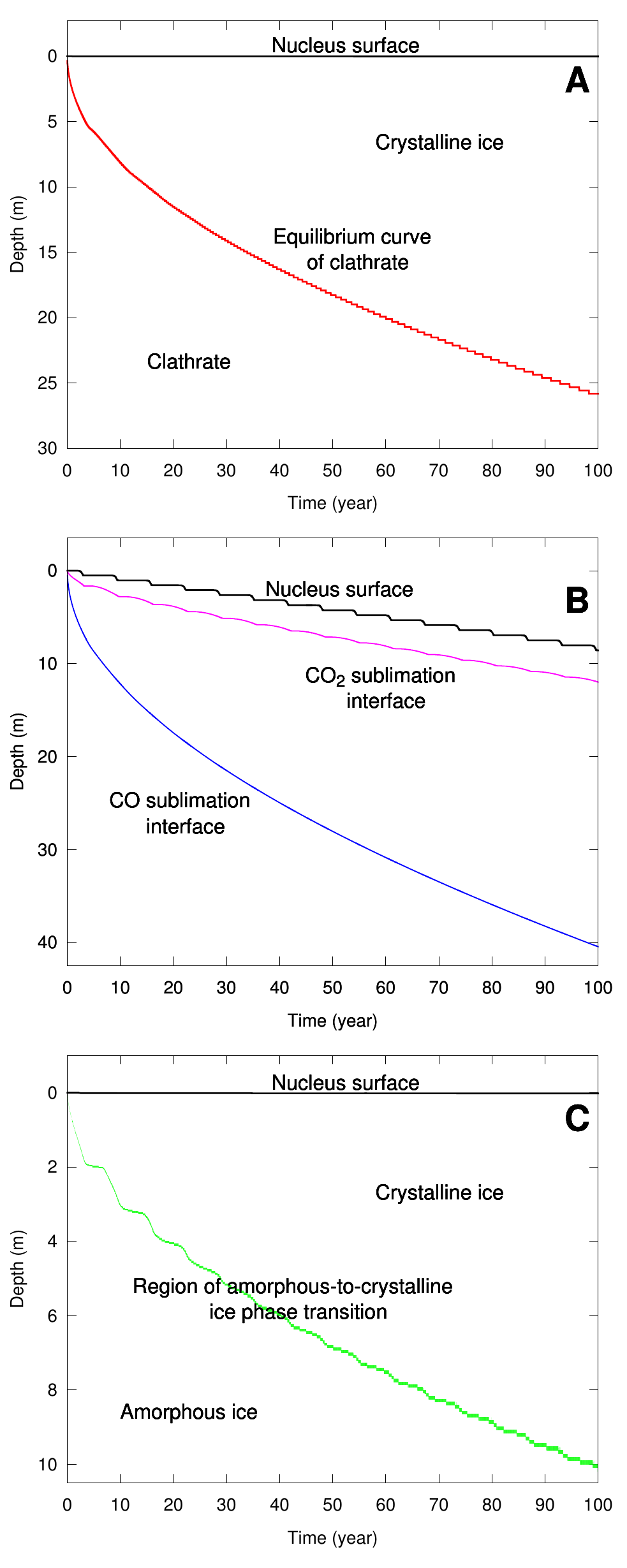}
\caption{Time evolution of 67P's nucleus stratigraphy at the location of the Seth$\_$01 region. Panel A: the initial volatile phase is formed from a mixture of CO and CO$_2$ trapped in clathrate. The red curve represents the boundary between the clathrate stability zone and the crystalline ice. Panel B: the initial material is a mixture of CO$_2$, CO and H$_2$O pure crystalline ices. The purple and blue curves represents the equilibrium curves of CO$_2$ and CO pure condensates, respectively. Panel C: the water ice matrix is initially amorphous, with a mixture of CO and CO$_2$ trapped within. The green area represents the subsurface region in which the amorphous-to-crystalline ice phase transition is occurring. Depth ``zero'' corresponds to the surface of the nucleus at the beginning of our computations. In cases A and C, a dust mantle of 1 to 10~cm forms at the surface, stopping the surface erosion.}
\label{fig:depth}
\end{center}
\end{figure}

Clathrates have the ability to trap up to one guest molecule for 5.75 or 5.67 water molecules, depending on the considered structure (Lunine \& Stevenson 1985; Mousis et al. 2010). Because of these large amount of volatiles potentially trapped in these structures, the explosive outgassing of clathrates has been invoked to explain the presence of chaotic terrains found on Earth, Mars, and many icy bodies of the outer solar system (Kargel et al. 2003). The breakdown of clathrates and subsequent release of volatiles towards the surface are possible candidates to destabilize the terrain of cometary surfaces, and create basins or pits such as those observed on the surface of 67P. In this section, we assume that the icy matrix is exclusively made of clathrates, with cages fully filled by CO and CO$_2$, with a CO/CO$_2$ ratio of 1.08 (see Table \ref{param}). The icy matrix is then composed of $\sim$82.6\%, 9\% and 8.4\% of H$_2$O, CO and CO$_2$, respectively. The results are shown in Figure \ref{fig:depth}A for a time evolution of the stratigraphy over a period of 100 years, i.e. $\sim$15.5 orbital periods of the comet. The local erosion of the nucleus surface is inhibited as a result of the formation of a dust mantle, whose thickness is in the 1 to 10~cm range, consistent with previous determinations by other models (Guilbert-Lepoutre et al. 2014 and references therein). 

The dust mantle causes a thermal lag preventing any heat wave to effectively result in the sublimation of water ice underneath, therefore maintaining the presence of crystalline water ice close to the surface. However, the clathrate stability region progressively moves away from the surface, resulting in the release of CO and CO$_2$ molecules throughout the porous matrix. A layer of about 26~m is destabilized over a 100~year period. A period of $\sim$800 years is required for the clathrate equilibrium curve to reach a depth of 200 m, which we choose as a reference for the depth of the Seth$\_$01 pit.

\subsection{Sublimation of crystalline ices}

The presence of a dust mantle inhibits the sublimation of crystalline ice, implying the location of the H$_2$O sublimation interface just beneath this layer, regardless of the considered epoch of 67P's orbital evolution. Therefore, in order to test the possibility that pits have been shaped by the sublimation of ices, the model requires the absence of dust in the matrix, or that dust grains are small enough to escape with gas. In other words, the model needs to be parameterized in such a way that no dust mantle can be built at the surface of the Seth$\_$01 region. Figure \ref{fig:depth}B represents the evolution of the subsurface, assuming that the icy matrix is made of crystalline ices of H$_2$O, CO and CO$_2$ with mole fractions identical to those used in the clathrate case. Here, the ablation due to water sublimation is of order of 0.55~m per orbit, implying that $\sim$2330 years of orbital evolution are needed to dig 200 m in depth. The sublimation of CO and CO$_2$ also contributes to the increase of porosity and embrittlement of the subsurface, but to a lesser extent than loss of H$_2$O would cause. After simulating 100 years of evolution, the CO and CO$_2$ sublimation interfaces lie $\sim$40.5 and 12.0 m beneath the surface of the Seth$\_$01 region.

\subsection{Amorphous-to-crystalline ice phase transition}

Because amorphous water ice has the ability to trap large amounts of volatiles, the amorphous-to-crystalline ice phase transition has been proposed to be at the origin of outbursts observed among comets for decades (Klinger 1980; Prialnik \& Bar-Nun 1987; Jewitt 2009; Guilbert-Lepoutre 2012). Here, the possibility of pits formation via this mechanism is evaluated and results are shown in Figure \ref{fig:depth}C. In this simulation, the initial icy matrix is exclusively made of porous amorphous water ice containing 10\% of CO and CO$_2$, residing on the ice walls due to Van Der Waals forces, with a CO/CO$_2$ ratio of 1.08 (see Table \ref{param}). Similarly to the clathrate case, a 1 to 10~cm dust mantle is progressively  built, quenching the water ice sublimation and stopping the surface erosion. With time, the phase transition front progressively moves toward the internal parts of the subsurface. After 100 years of orbital evolution, the limit between amorphous and crystalline ice is located at a $\sim$10 m depth. A period of $\sim$2000 years is needed for the phase transition to occur at a depth of 200~m.

\section{Discussion}

{Our simulations show that clathrate destabilization and amorphous ice crystallization can occur at depths corresponding to those of the pits observed in the Seth$\_$01 region. In contrast with the recent estimate of Vincent et al. (2015), we find that sublimation of crystalline ice (mainly H$_2$O ice) down to such depths} is possible only in the absence of a dust mantle, which requires the presence of dust grains in the matrix small enough to be dragged out by gas from the pores. 
The porosity of the dust mantle plays an important role in the process. Depending on the local porosity, the thermal destabilization of clathrates or amorphous ice in subsurface layers can be induced regularly (high porosity where gas molecules can be released easily), or episodically through eruptive processes (lower porosity allowing for gas pressure build-up underneath the dust mantle). To a larger extent, a sealed dust mantle would favor episodic and disruptive outgassing as a result of an increasing gas pressure in the pores, while a high porosity should allow the formation of large voids in the subsurface due to the continuous escape of volatiles, a process consistent with the formation of sinkholes (Belton and Melosh 2009). {Our results are therefore consistent with both pits formation via sinkholes (Belton and Melosh 2009) or directly due to violent outbursts (Belton et al 2013; Thomas et al. 2013)}.

The thermal processing of 67P's subsurface can occur at depths typical of those estimated for the observed pits, in a timescale ($\sim$1000--2300 yr) shorter than 67P's lifetime of $\sim$7000 yr estimated by Groussin et al. (2007) in the comet's activity zone in the inner solar system. However, this depth is correlated with both thermo-physical properties, and the orbital evolution of the comet in the inner solar system. For example, a recent study by Maquet (2015) suggests that 67P's perihelion distance was much further ($>$ 2.7 AU) than the current value (1.3 AU) prior the close encounter with Jupiter in 1959. If this estimate is correct, the illumination conditions studied with our model are not valid throughout the 1000--2300~yr period which have been found to dig the observed pits. To form these pits after 1959, a local high value of the thermal inertia would be required. For example, we performed the same study with a thermal inertia of 1000 W K$^{-1}$m$^{-2}$s$^{1/2}$. For the clathrate case, the stability region is located $\sim$200~m deep after 50 years of cometary evolution. For the amorphous ice case, the phase transition front is located $\sim$90~m deep. 

{In order to investigate the formation conditions of pits under different illumination conditions, we have considered the extreme case of the Ma'at\_04 region localized on 67P's southern hemisphere. This region experiences full illumination at perihelion, in contrast with the Seth\_01 zone which is essentially in polar night during this period. Using the same input parameters, our computations show that the cases of clathrate and amorphous ice destabilizations behave similarly to those investigated for the Seth\_01 region, with similar depths reached by the different interfaces over a period of 100 years. This results from the fact that the thermal insulation created by the dust mantle remains very efficient regardless of the illumination conditions. However, with a strong water ice ablation ($\sim$200~m after 100 years of cometary evolution), the crystalline ice case is consistent with pit formation in short timescales in the Ma'at\_04 region. If the mechanism of crystalline ice sublimation is dominant, the pits localized in the Ma'at\_04 region are expected to be younger than those located in less illuminated regions. If one single mechanism is responsible for the formation of pits, then the sublimation of ice requires different dust properties across the surface.}

{However, our study does not account for the influence of shadowing and self-heating, which could strongly affect the local thermal balance and energy available for heating the subsurface. Self-heating can become efficient when pits are deep, and when parts of the walls receive direct solar illumination. In contrast, shadowing may locally decrease the floor's temperature and the amount of heat transferred to the subsurface. When effective, thermal involvement of pit walls likely make bowls rather than cylinders (Byrne \& Ingersoll 2003). This process can speed up the digging of an initially smaller pit, thus requiring a lower thermal inertia, more consistent with the observations described in Leyrat et al. (2015).} In consequence, based on our computations, it is difficult to determine whether the observed pits were formed since 1959 when the comet became potentially more active, or before that when the energy input was lower at the surface compared to what we have studied here (although still adequate for forming pits during the estimated lifetime of 67P in the inner solar system). However, an observational test would be to see whether the pit significantly deepened or expanded during the Rosetta period of close observations, because one perihelion passage is a significant part of total time spent near perihelion since 1959.  We might expect to see some very visible changes. If instead the comet has spent several hundred perihelion passages very near the sun over the past couple thousand years, changes might not be evident.  

{In addition, our study does not address the shape of pits, which depends on the unknown mechanical layering of material in the subsurface. Regardless of the process, the formation of pits requires that the surface and subsurface are locally heterogeneous in composition and/or in thermo-physical properties, otherwise the overall surface would be much smoother, with similar processes occurring everywhere at the same time and the same rate.} The case of ice sublimation for instance requires that local dust grains may be small enough to avoid dust mantling, when the pits' surroundings are covered with a dust mantle. {Pits formed in different regions may require varying dust properties. Illumination conditions similar to those studied in pit forming regions occur at other areas of the surface, where pits are not observed. This implies variable physical or chemical properties across the surface and subsurface, and hence varying susceptibilities to pit formation.} We can therefore firmly conclude that the subsurface of 67P is not uniform at a spatial scale of $\sim$100--200~m. 

{Ancient or current thermal processing, due to cometary activity, could be at the origin of this heterogeneity. 67P, as with other comets mapped by spacecrafts, shows movements of materials across its surface (Groussin et al. 2015). The local albedo, as well as the thickness and thermal properties of the dust mantle, could then vary according to dust deposition. Thermal shadowing could also play an important role (Guilbert-Lepoutre \& Jewitt 2011). Interestingly, thermal modeling of homogeneous substrate with variable surficial roughness may explain pit growth on 67P, as shown in the case of the CO$_2$ ice cap on Mars (Byrne et al. 2015).} Alternatively, it may be as well an argument in favor of early collisions, during which a nucleus accretes various materials (Belton et al. 2007). {If the pit size is representative of that of accreted planetesimals,} the spatial scale obtained from the presence of pits on 67P is consistent with those previously found for other comets, ranging from 50--100~m obtained from the observations of split comets (Guilbert-Lepoutre et al. 2015 and references therein) to 400~m obtained from the layering observed on comet 9P/Tempel~1 (Belton et al. 2007).\\

\acknowledgements
We thank an anonymous Referee for his constructive comments thats helped us strengthen our manuscript. O.M. acknowledges support from CNES. This work has been partly carried out thanks to the support of the A*MIDEX project (n\textsuperscript{o} ANR-11-IDEX-0001-02) funded by the ``Investissements d'Avenir'' French Government program, managed by the French National Research Agency (ANR).

\clearpage

\clearpage

\begin{thebibliography}{}

\bibitem[Belton et al. (2007)]{Bel07} Belton, M.J.S., Thomas, P., Veverka, J., et al.\ 2007, Icarus, 187, 332

\bibitem[Belton et al.(2013)]{2013Icar..222..477B} Belton, M.~J.~S., Thomas, P., Carcich, B., et al.\ 2013, Icarus, 222, 477 

\bibitem[Belton \& Melosh(2009)]{2009Icar..200..280B} Belton, M.~J.~S., \& Melosh, J.\ 2009, Icarus, 200, 280 

\bibitem[Brownlee et al.(2004)]{2004Sci...304.1764B} Brownlee, D.~E., Horz, F., Newburn, R.~L., et al.\ 2004, Science, 304, 1764 

\bibitem[Byrne et al.(2015)]{2015LPI....46.1657B} Byrne, S., Hayne, P.~O., Becerra, P., \& HiRISE Team 2015, Lunar and Planetary Science Conference, 46, 1657 

\bibitem[Byrne \& Ingersoll(2003)]{2003GeoRL..30.1696B} Byrne, S., \& Ingersoll, A.~P.\ 2003, \grl, 30, 1696

\bibitem[de Sanctis et al.(2005)]{DeS05} de Sanctis, M. C., Capria, M. T.; \& Coradini, A., 2005. \aap, 444, 605

\bibitem[Fornasier et al.(2015)]{2015.1399G} Fornasier, S., Hasselmann, P. H., \& Barucci, M. A., et al.\ 2015, \aap, in press (http://arxiv.org/pdf/1505.06888v1.pdf)

\bibitem[Groussin et al.(2015)]{2MNRAS.376.1399G} Groussin, O., Sierks, O., Barbieri, C., et al. \aap, in press (http://arxiv.org/pdf/1509.02794.pdf)

\bibitem[Groussin et al.(2007)]{2007MNRAS.376.1399G} Groussin, O., Hahn, G., Lamy, P.~L., et al.\ 2007, \mnras, 376, 1399 

\bibitem[Guilbert-Lepoutre et al.(2015)]{2015SSRv..tmp...23G} Guilbert-Lepoutre, A., Besse, S., Mousis, O., et al.\ 2015, \ssr, 23 

\bibitem[Guilbert-Lepoutre et al.(2014)]{2014A&A...567L...2G} Guilbert-Lepoutre, A., Schulz, R., Rozek, A., et al.\ 2014, \aap, 567, L2

\bibitem[Guilbert-Lepoutre(2012)]{2012AJ....144...97G} Guilbert-Lepoutre, A.\ 2012, \aj, 144, 97 

\bibitem[Guilbert-Lepoutre \& Jewitt(2011)]{2011ApJ...743...31G} Guilbert-Lepoutre, A., \& Jewitt, D.\ 2011, \apj, 743, 31

\bibitem[Jewitt(2009)]{2009AJ....137.4296J} Jewitt, D.\ 2009, \aj, 137, 4296 

\bibitem[Jorda et al.(2014)]{2014AGUFM.P41C3943J} Jorda, L., Gaskell, R.~W., Hviid, S.~F., et al.\ 2014, AGU Fall Meeting Abstracts, 3943 

\bibitem[Kargel et al.(2003)]{2003EAEJA....14252K} Kargel, J.~S., Prieto-Ballesteros, O., \& Tanaka, K.~L.\ 2003, EGS - AGU - EUG Joint Assembly, 14252

\bibitem[Keller et al.(1988)]{1988Natur.331..227K} Keller, H.~U., Kramm, R., \& Thomas, N.\ 1988, \nat, 331, 227 

\bibitem[Klinger(1980)]{1980Sci...209..271K} Klinger, J.\ 1980, Science, 209, 271 

\bibitem[Kossacki \& Szutowicz(2006)]{Kos06} Kossacki, K. J. \& Szutowicz, S., 2006, P\&SS, 54, 15

\bibitem[LeRoy et al.(2015)]{L15} Le Roy, L., Altwegg, K., Balsiger, et al.\ 2015, \aap, in press

\bibitem[Leyrat et al. (2015))]{2015Natur.331..227K} Leyrat, C., Erard, S., Capaccioni, F., et al. 2015, EGU General Assembly Conference Abstracts, 17, 9767 

\bibitem[Lunine \& Stevenson(1985)]{1985ApJS...58..493L} Lunine, J.~I., \& Stevenson, D.~J.\ 1985, \apjs, 58, 493

\bibitem[Maquet(2015)]{2015A&A...579A..78M} Maquet, L.\ 2015, \aap, 579, A78 

\bibitem[Marboeuf et al.(2012)]{2012A&A...542A..82M} Marboeuf, U., Schmitt, B., Petit, J.-M., Mousis, O., \& Fray, N.\ 2012, \aap, 542, A82 

\bibitem[Mottola et al.(2014)]{2014A&A...569L...2M} Mottola, S., Lowry, S., Snodgrass, C., et al.\ 2014, \aap, 569, L2 

\bibitem[Mousis et al.(2010)]{2010FaDi..147..509M} Mousis, O., Lunine, J.~I., Picaud, S., \& Cordier, D.\ 2010, Faraday Discussions, 147, 509 

\bibitem[Preusker et al.(2015)]{2015..147..509M} Preusker, F., Scholten, F., Matz, K.-D., et al.\ 2015, \aap, in press

\bibitem[Prialnik \& Bar-Nun(1987)]{1987ApJ...313..893P} Prialnik, D., \& Bar-Nun, A.\ 1987, \apj, 313, 893

\bibitem[Rosenberg \& Prialnik(2009)]{Ros09} Rosenberg, E.D. \& Prialnik, D., 2009. Icarus, 201, 740

\bibitem[Rotundi et al.(2014)]{2014M&PS...49..550R} Rotundi, A., Rietmeijer, F.~J.~M., Ferrari, M., et al.\ 2014, Meteoritics and Planetary Science, 49, 550 

\bibitem[Soderblom et al.(2002)]{2002Sci...296.1087S} Soderblom, L.~A., Becker, T.~L., Bennett, G., et al.\ 2002, Science, 296, 1087 

\bibitem[Thomas et al.(2015)]{2015Sci...347.0440T} Thomas, N., Sierks, H., Barbieri, C., et al.\ 2015, Science, 347, aaa0440 

\bibitem[Thomas et al.(2013)]{2013Icar..222..550T} Thomas, P.~C., A'Hearn, M.~F., Veverka, J., et al.\ 2013, Icarus, 222, 550

\bibitem[Tubiana et al.(2015)]{2015A&A...573A..62T} Tubiana, C., Snodgrass, C., Bertini, I., et al.\ 2015, \aap, 573, A62 

\bibitem[Vincent et al.(2015)]{2015Icar..222..550T} Vincent, J.-B., Bodewits, D., Besse, S., et al.\ 2015, Nature, 523, 63

\end{thebibliography}
\end{document}